\documentclass[floatfix,aps,twocolumn,amsmath,amssymb,superscriptaddress,prl]{revtex4-2}
\usepackage{bm,bbm}
\usepackage{amssymb,amsfonts,amsmath}
\usepackage{graphicx}
\usepackage{xcolor}
\usepackage{ulem}
\usepackage{SIunits}
\usepackage[utf8]{inputenc}
\usepackage[english]{babel}
\usepackage{hyperref}
\usepackage{ulem}
\usepackage{braket}
\usepackage{bpchem}
\graphicspath{ {./Bilder/} }
\usepackage{tikz}
\usepackage{pdfpages}
\usepackage{url}

\makeatletter
\AtBeginDocument{\let\LS@rot\@undefined}
\makeatother

\newcommand{\abs}[1]{\left| #1 \right|}

\newcommand{\ex}[1]{\left\langle  #1 \right\rangle}

\newcommand{\RM}[1]{\MakeUppercase{\romannumeral #1}}
\newcommand{\rs}[1]{\rm{\scriptscriptstyle #1}}

\begin{document}

\title{
Quantum fluctuations in one-dimensional supersolids
}
\author{Chris Bühler}
\affiliation{
  Institute for Theoretical Physics \RM3 and Center for Integrated Quantum Science and Technology, University of Stuttgart, DE-70550 Stuttgart, Germany
}
\author{Tobias Ilg} 
\affiliation{
  Institute for Theoretical Physics \RM3 and Center for Integrated Quantum Science and Technology, 
  University of Stuttgart, DE-70550 Stuttgart, Germany
}
\author{Hans Peter Büchler}
\affiliation{
  Institute for Theoretical Physics \RM3 and Center for Integrated Quantum Science and Technology, University of Stuttgart, DE-70550 Stuttgart, Germany
}
\date{\today}

\begin{abstract}
In one-dimension, quantum fluctuations prevent the appearance of long-range order in a  supersolid,
and only quasi long-range order can survive. We derive this quantum critical behavior  and study its influence on
the superfluid response and properties of the solid.  The analysis is based on an effective low-energy description accounting
for the two coupled Goldstone modes. We find that the quantum phase transition from the superfluid to the supersolid is shifted 
by quantum fluctuations from its mean-field prediction. However, for current experimental parameters with dipolar atomic gases, 
this shift is not observable and the transition appears to be mean-field like.
\end{abstract}
\maketitle

A remarkable property of quantum fluctuations is that they strongly influence spontaneous symmetry breaking 
in one-dimensional systems.  Especially, it is well established that one-dimensional superfluids
only exhibit quasi long-range order with a characteristic algebraic decay \cite{Schwartz1977,Haldane,Giamarchi:743140,Petrov2000}.
Nevertheless, the latter can support a superfluid flow across a weak impurity
\cite{Kagan2000,Buechler2001}.  Recent experiments with weakly interacting dipolar Bose gases  have 
observed the appearance of a supersolid phase in elongated one-dimensional geometries \cite{Tanzi2019a,Boettcher2019,Chomaz2019}. Such
a supersolid state breaks the translational symmetry giving rise to a solid structure as well as 
the $U(1)$ symmetry for the superfluid \cite{Boninsegni2012}. The experimental observations are in good agreement 
with mean-field theory  within the  extended  Gross-Pitaevskii  formalism \cite{Boettcher2020,Chomaz2022}, which also predicts 
a second order  quantum phase transition from the superfluid to the supersolid  in the thermodynamic limit \cite{Roccuzzo2019,Ilg2022}.
In this letter, we study the influence of 
quantum fluctuations on such one-dimensional supersolid phases  in the thermodynamic limit.

The effect of quantum fluctuations has been intensely studied for superconducting thin wires:  in addition to the appearance of quasi long-range order, 
the quantum nucleation of phase slips provide dissipation even in the superconducting phase at zero temperature and give rise to an algebraic current-voltage characteristic \cite{Zaikin1997,Golubev2001,Kashurnikov1996,Hekking1997,Buechler2004,Meidan2007}.  Such an algebraic current-voltage characteristic still gives rise to a superconducting phase 
as the linear resistivity vanishes. However, quantum fluctuations  can eventually  drive a quantum phase transition from the superconductor to a state with finite resistivity 
 for increasing influence of the phase fluctuations \cite{Zaikin1997,Golubev2001,Kashurnikov1996,Hekking1997,Buechler2004,Meidan2007}. For a Galilei invariant superfluid, quantum nucleation of phase slips can only appear at an impurity,  
 but  similarly  give rise to dissipation and eventually drive a quantum phase transition \cite{Kagan2000,Buechler2001,DErrico2017,Polkovnikov2005}.   
 In analogy, the phonon mode in a solid gives rise to a similar effective low-energy description, and therefore  quantum fluctuations also strongly
 affect  the properties of a solid \cite{Dalmonte2010}.
 This opens the question how quantum fluctuations influence the properties of a supersolid -- a state, which combines
 the characteristics of a superfluid and a solid \cite{Leggett1970,Boninsegni2012}. While the search of such supersolid states focused on systems with an 
 averaged particle number per lattice site close to one \cite{Balibar2010,Boninsegni2012,Chan2013}, the recent experiments with Dysprosium atoms
 are in a complementary regime with a large amount of particles per lattice site  \cite{Tanzi2019a,Boettcher2019,Chomaz2019}. The effective low-energy description of such a supersolid has been derived  based on a local mean-field theory \cite{Joss_Supersolids,Josserand2007,Yoo_Dorsey}.

In this letter, we analyze the influence of quantum fluctuations on the quantum  phase transition from the superfluid to the supersolid and
 study the quasi long-range order in the supersolid phase in the parameter regime realized by recent experiments with Dysprosium atoms. 
 The analysis is based on the effective low-energy theory for a supersolid
with many particles within a lattice site \cite{Joss_Supersolids,Josserand2007,Yoo_Dorsey}. The superfluid is defined by the ability of the system 
to sustain a dissipationless particle flow across a weak impurity, i.e., absence of a linear relation between flow and pressure \cite{Kagan2000,Buechler2001}. In analogy the solid character is defined by the ability of the system
to drag the solid structure with a moving impurity. 
We find that the quantum phase
transition from the superfluid to the supersolid is shifted:  the formation of a local solid structure takes place first at the mean-field transition, while the
supersolid phase only appears for sufficient correlations between these local solid structures.
 However, for current experimental setups with dipolar quantum gases, this shift is not observable
and therefore, the quantum phase transition is extremely well described by mean-field theory.
Finally, we study the disappearance of the supersolid phase  for increasing correlations.

We start with the effective low-energy description of a one-dimensional supersolid consisting of  weakly interacting bosons with a large number of atoms per lattice site.
Then,  the bosonic field operator can be written as 
$\psi({\bm x}) = \sqrt{\rho({\bm x})}\, e^{i \varphi({\bm x})}$ 
with the phase field $\varphi$ and the density field
\begin{align}
\rho({\bm x}) &= \left[n+ \delta n({\bm x})\right]
f\left( x - \frac{d}{2\pi} u({\bm x})\right), \label{eq:density}
\end{align}
while we introduced the notation ${\bm x}= (x,t)$ for the space-time coordinate. 
Here,  $f(x)= f(x\!+\!d)$ is a periodic function with period $d$ and normalized to $\int_0^d dx\, f(x)/d=1$; 
it accounts for the local formation of a solid-like structure by droplets, 
while the displacement  field $u({\bm x})$ allows for fluctuations in the position of these droplets. 
Note, that $n$ denotes the averaged density with $n d$ particles within each droplet, while  $\delta n({\bm x})$ describes local density fluctuations. 
The low-energy behavior is then captured by the effective Lagrangian for the slowly varying fields $\varphi$, $\delta n$ and $u$ \cite{Joss_Supersolids,Josserand2007,Yoo_Dorsey}, 
\begin{align}
	\mathcal{L}=&-\hbar\delta n\partial_{t}\varphi- \frac{\kappa}{2}\left(\delta n\right)^2 - \frac{\lambda'}{2}\left(\partial_{x}u\right)^2 - \xi' \delta n \partial_{x}u \nonumber\\
	&+\frac{\hbar^2 n}{2m}\left[ \frac{n_{\rs{L}}}{n}\left(\frac{m d}{2\pi \hbar}\partial_{t}u-\partial_{x}\varphi\right)^2- \left(\partial_{x}\varphi\right)^2 \right]. 
	\label{eq:lagrangian}
\end{align}
The second line corresponds to the kinetic energy, 
where the term $\partial_t u$ accounts for the velocity of the droplet at position $x$, 
while the superfluid exhibits a reduced superfluid density 
$n_{\rs{s}}\equiv n-n_{\rs{L}}$ due to the formation of a local solid-like structure \cite{Leggett1970}.
Furthermore, the first line includes the conventional coupling between the phase field and the density in a superfluid as well as an
expansion of the interaction energy to second order  in the slowly varying fields with parameters $\kappa$, $\lambda'$ and $\xi'$. 
Note, that these parameters can be conveniently derived within mean-field theory \cite{Ilg2022,supplement}, 
and the stability of the system naturally requires $\kappa \lambda' - (\xi')^2>0$.
In the weakly interacting regime, 
we also require 
$\hbar^2n/(m\kappa)\gg1$. i.e., the kinetic energy of the  superfluid,  
is much larger than the interaction energy.

The Lagrangian in Eq. \eqref{eq:lagrangian} describes a strong coupling between the Bogoliubov mode of the superfluid and the phonon mode of a solid, and it gives rise to two linear sound modes accounting for the two broken symmetries.
Furthermore, we find the current conservation $\partial_t\delta n= -\partial_x (j_{\rs{s}} + j_{\rs{n}})$ 
with the normal and superfluid current $j_{\rs{n}}=(n_{\rs{L}}d/2 \pi)\partial_t u$ and $j_{\rs{s}}=(\hbar n_{\rs{s}}/m)\partial_x\varphi$.
For $n_{\rs{L}}/n\rightarrow 0$ the solid structure disappears. 
Consequently, $\xi'\rightarrow 0$ and $\lambda'\rightarrow 0$,  
and we recover the effective low-energy description of a superfluid.
For $n_{\rs{L}}/n\rightarrow 1$,  we obtain the theory of phonons in a solid with the compressibility 
$(n^2 \kappa+ 4\pi^2\lambda'/d^2 - 4\pi n \xi'/d)/m$.

In the following, it is convenient to switch to a Hamiltonian description of the low-energy quantum theory,
\begin{align}\label{eq:Hamiltonian_Matrix_Form}
 H= \frac{\hbar}{2 \pi}\int dx  &\left[ v_{J} 
 \begin{pmatrix}
 \partial_{x}\varphi\\
 \partial_{x}w
 \end{pmatrix} {\bf M}_{J} \begin{pmatrix}
 \partial_{x}\varphi\\
 \partial_{x}w
 \end{pmatrix}	\right. \nonumber\\
 &+ \left.  v_{N} 
 \begin{pmatrix}
 \partial_{x}\vartheta\\
 \partial_{x}u
 \end{pmatrix} {\bf M}_{N} \begin{pmatrix}
 \partial_{x}\vartheta\\
 \partial_{x}u
 \end{pmatrix}\right],
\end{align}
where $-\hbar \partial_{x} \vartheta/\pi$ and $-\hbar \partial_{x} w/\pi$ denote the conjugate variables to $\varphi$ and $u$, respectively.
We have introduced the two velocities $v_{J} = \hbar \pi n/m$ and $v_{N} = \kappa/\pi \hbar$, 
with $v_N/v_J\ll1$ in the weakly interacting regime.
The matrices ${\bf{M}}_J$ and ${\bf{M}}_N$ take the form
\begin{equation}
 {\bf M}_{J}= \left(\begin{array}{c c}
 1 & -\beta\\
- \beta&  \beta^2/\gamma
 \end{array}\right), 
 \quad 
 {\bf M}_{N}=\left(\begin{array}{c c}
 1 & \xi\\
\xi& \lambda
 \end{array}\right), 
\end{equation}
with $\beta = 2/n d$,  $\gamma= n_{\rs{L}}/n$ 
and the dimensionless parameters 
$\lambda= \lambda'\pi^2/\kappa$ and $\xi = \xi'\pi/\kappa$. 
Since $\gamma\leq 1$, ${\bf M}_{J}$ is positive semi definite. 
The stability in the thermodynamic limit requires ${\bf M}_{N}$ to be positive semi-definite as well, i.e. $\lambda - \xi^2 \geq 0$. 
Being conjugate variables, the canonical commutation relations read
\begin{equation}
	\left[\partial_{x} \vartheta({\bm x}), \varphi({\bm y})\right] = i \pi \delta({\bm x}-{\bm y}) = \left[\partial_{x}u({\bm x}), w({\bm y})\right].
	\label{eq:commutator}
\end{equation}
It is possible to diagonalize this Hamiltonian into two uncoupled sound modes by the transformation, 
\begin{align}
    \begin{pmatrix}
     \phi_+\\
    \phi_-
     \end{pmatrix}= {\bf Q}  
     \begin{pmatrix}
     \varphi\\
     w
     \end{pmatrix},
     \quad
    \begin{pmatrix}
    \theta_+\\
    \theta_-
    \end{pmatrix}
    = \left({\bf Q}^{-1}\right)^{T}  
    \begin{pmatrix}
    \vartheta \\
    u
    \end{pmatrix}.
\end{align}
The construction of this canonical transformation is presented in 
\cite{supplement}
and we obtain the Hamiltonian 
\begin{equation}
 H= \frac{\hbar}{2 \pi} \int dx \sum_{\sigma\in\{+,-\}} v_{\sigma}\left[ \left(\partial_{x}\phi_\sigma\right)^2  + \left(\partial_{x}\theta_\sigma\right)^2 \right]
 \label{eq:hamilton}
\end{equation}
with the two sound velocities
\begin{equation}
    v_{\pm}^2 = \frac{v_{J} v_{N}}{2 }\left[ \alpha \pm \sqrt{\alpha^2-4\beta^2(\lambda-\xi^2)(1\!-\!\gamma)/\gamma} \right]
\end{equation}
and $\alpha=1-2\xi\beta+\beta^2\lambda/\gamma$.
The Hamiltonian  in Eq.~(\ref{eq:hamilton}) allows us to derive the behavior of correlation functions at long distances \cite{Haldane}. 
We obtain quasi long-range off-diagonal order for the superfluid
as well as quasi long-range diagonal order for the solid, 
\begin{eqnarray}
    \ex{\psi(x)\psi^\dag(0)} &=  &n \left( \frac{\zeta}{|x|}\right)^{A/2} \\
      \ex{\rho(x)\rho(0)}\! -\!n^2&= & - \frac{C}{2 \pi^2 \abs{ x}^2} 
    + \eta \cos\frac{2\pi x}{d}\left(\frac{\zeta}{\abs{x}}\right)^{ B/2} . \nonumber
\end{eqnarray}
Here, we introduced the non-universal parameter $\eta$, and the short-distance cut-off $\zeta$, while the algebraic decay
is determined by the canonical transformation via  $A = (({\bf Q}^{-1})_{11})^2 + (({\bf Q}^{-1})_{12})^2$, 
$B = ({\bf Q}_{12})^2 + ({\bf Q}_{22})^2 $, and $ C=({\bf Q}_{11})^2 + ({\bf Q}_{21})^2$. 

\begin{figure}
    \includegraphics[width=\columnwidth]{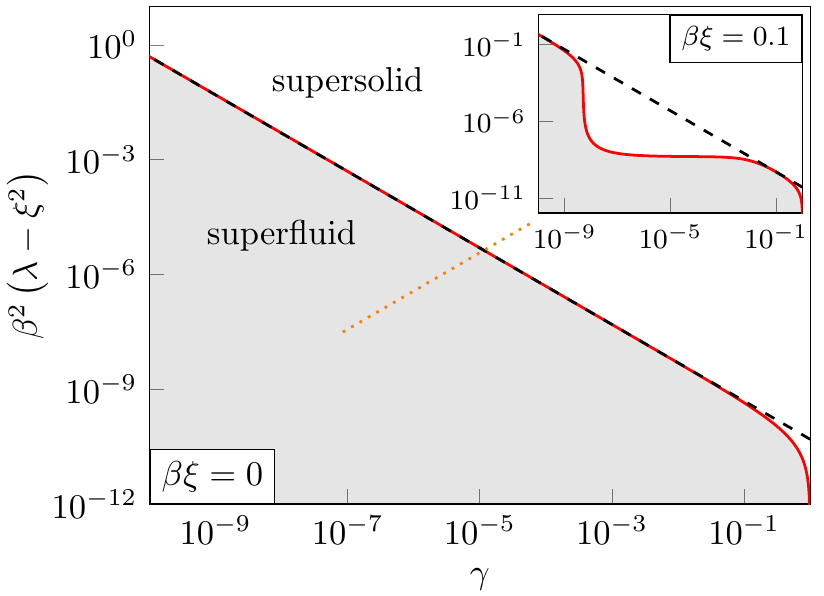}
    \caption{
    Critical line $B=4$ (red solid line) as a function of $\gamma$ and $\beta^2(\lambda-\xi^2)$ at $\beta\xi=0$ for fixed $v_{J}/v_{N}\approx1.0\times10^{7}$ and $\beta\approx 2.1\times 10^{-4}$. 
    The black dashed line shows the asymptotic behavior of $B$ for $\gamma\rightarrow 0$. 
    In the gray shaded region $B>4$ and the system does not feel the impurity (superfluid), 
    while in the white region $B<4$ the perturbation becomes relevant (supersolid).
    The orange dotted line shows the path across the phase transition for experimentally realistic parameters.
    In the inset, we fix $\beta\xi=0.1$ while $v_J/v_N$ and $\beta$ remain unchanged.
}
    \label{fig:contour}
\end{figure}

In the following, we study the quantum phase transitions in the system. 
The characteristic property of a superfluid is that it can sustain a superfluid flow, 
while in a solid a moving localized impurity  can drag  the solid structure along; 
a supersolid exhibits both of these properties, i.e., 
it can sustain a superfluid flow while a moving impurity drags the solid  structure along.  
These conditions provide critical values for the algebraic correlations above and will be studied in the following. 

We start with the parameters in the superfluid close to the formation of a solid-like structure,  i.e., $\gamma \ll 1$
with $A \sim \sqrt{v_N/v_J} \ll 1$; this condition is sufficient to sustain a superfluid flow, see below. 
Therefore, we first study the transition into the supersolid for increasing  $\gamma$, i.e., stronger local solid-like structure. 
A local impurity at position $x_0$ is described by an external potential $V_{I} \approx g \delta(x-x_0)$, 
and provides a  contribution to the low-energy Hamiltonian 
\begin{equation}
 H_{\rs I}= \!\!\int dx \rho(x) V_{I}(x) \sim   g_u \cos(u(x_0) + 2 \pi x_0 /d  )  \label{eq:uterm},
\end{equation}
where we expanded the local solid structure $f(x)$  in Eq.~(\ref{eq:density}) into a Fourier series.
Note that the impurity can provide additional terms when taking the discrete nature of particles into account \cite{Haldane}, but these do not become relevant before superfluidity is lost (see below).
The low-energy description then reduces to a coupled boundary sine-Gordon model \cite{KaneFisher,Ghoshal1994,Fendley1994,Fendley1995,Chudzinski_Gabay_Giamarchi,Kundu_2021}.
The term in Eq.~(\ref{eq:uterm}) is irrelevant for $B>4$ 
and therefore the system does not feel the presence of the impurity in the low-energy regime. In turn, the term  becomes relevant for  $B<4$ and pins $u(x_0)$ to the minimum of the cosine.  Varying the position $x_0$ of the impurity now results in a change in $u$, 
which shifts the local solid structure of the system with the impurity.  
Hence, the system exhibits a solid character for $B<4$.
In our dimensionless units $B$ is given by
\begin{eqnarray}\label{eq:B_expanded}
    B & = & \beta^2\frac{v_J}{\left(v_+ + v_-\right)}\left[\frac{1}{\gamma}+\sqrt{\frac{1-\gamma}{\gamma \beta^2\left(\lambda-\xi^2\right)}} \: \:\right] 
    \\
    &\sim &\beta^2\sqrt{\frac{v_J}{v_N}} \frac{1}{\sqrt{\gamma\beta^2\lambda}} \hspace{20pt} \mbox{for $\gamma \rightarrow 0$}  \nonumber \,.
\end{eqnarray}
In Fig.~\ref{fig:contour} the critical line $B=4$ of the quantum phase transition separating the superfluid from the supersolid is shown
for different values of  $\xi$.
The parameters  $v_{J}/v_{N}= 1.0\times 10^{7}$ and $\beta=  2.1\times 10^{-4}$ are fixed to realistic values derived within
mean-field theory for an experimentally realistic setup (see below). 
The transition always takes place at a finite and non-vanishing  value of $\gamma$, 
i.e., the mean-field transition appearing at $\gamma=0$ is separated by quantum fluctuations from the true quantum phase transition into the supersolid.  

It is important to note, that the local formation of droplets can act as a source for the nucleation of quantum phase slips. Even in the superfluid as well as in the supersolid, such phase slips will give rise to a small dissipation and an algebraic behavior between the pressure difference $\Delta P$ for sustaining the  particle 
current $I$ with $\Delta P \sim I^{2/A - 2}$; this behavior is in analogy to thin superconducting wires \cite{Zaikin1997}. 

\begin{figure}[tb]
    \includegraphics{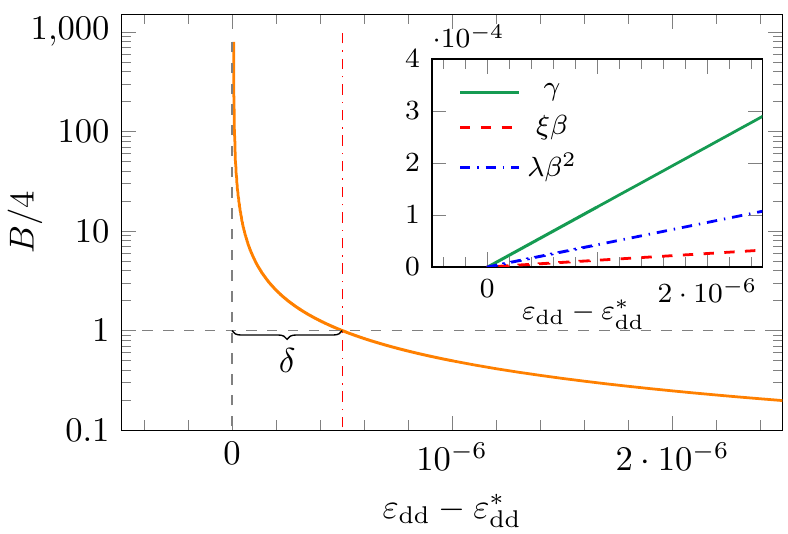}
    \caption{
    Superfluid to supersolid transition for increasing dipolar strength.
    We show $B/4$ as a function of $\varepsilon_\text{dd}-\varepsilon^*_\text{dd}$, where $\varepsilon^*_\text{dd}=1.34$ is the mean-field critical point.
    The system transitions to the supersolid at $B/4=1$ (red vertical line), 
    which is shifted compared to the mean-field transition at $\gamma=0$ (gray vertical line). 
    The inset shows the system parameters $\gamma$, $\beta\xi$ and $\beta^2\lambda$ used to calculate $B/4$ as a function of $\varepsilon_\text{dd}$
    for a harmonic transverse trapping with oscillator length $l_\perp=200a_s$ and one-dimensional density $n=11.931/a_s$.
    }
    \label{fig:transition}
\end{figure}

In the following, we analyze these phenomena for an experimentally
realistic setup based on Dysprosium atoms. The mean-field analysis for such a weakly 
interacting dipolar quantum gas in a one-dimensional geometry
predicts a second order quantum phase transition from the superfluid to a supersolid phase \cite{Ilg2022}.
For a transverse  harmonic trapping  with  length $l_\perp= 200 a_s$ and a density 
$n \approx 11.931 /a_s$, the mean-field formation of droplets takes place at $\varepsilon^*_{\text{dd}}=1.34$; here, $a_s$ denotes the $s$-wave scattering length and $\varepsilon_\text{dd}$ the relative dipolar interaction strength. 
For such a set of parameters, we obtain $v_{J}/v_{N} \approx 1.0\times 10^{7} $ and $\beta \approx 2.1\times 10^{-4} $, as well as the parameters $\gamma$, $\xi \beta$ and $\lambda \beta^2$ \cite{supplement}, see inset of Fig.~\ref{fig:transition}.
As predicted above, we find that $\xi$ and $\lambda$ vanish for $\gamma \rightarrow 0$. 
Then, the behavior of the parameter $B/4$ is shown in Fig.~\ref{fig:transition} as a function of  $\varepsilon_\text{dd}-\varepsilon^*_\text{dd}$. 
Indeed, we find that the quantum phase transition to the supersolid is shifted by quantum flucations from the mean-field prediction at $\varepsilon^*_{\text{dd}}=1.34$.
However, this region is extremely small with $\delta\approx 5\times 10^{-7}$ and beyond current control on the experimental parameters.
Therefore, the transition from the superfluid into the supersolid phase is extremely well described by mean-field theory for current experimental setups. 

Finally, we can also analyze the quantum phase transition from the supersolid into a solid, which appears for $\gamma \rightarrow 1$.
One can understand this transition as in this regime, 
the different droplets of the solid structure are only connected by a very weak superfluid link and essentially give rise to a Josephson junction between each droplet. 
Note that such a Josephson junction can only support a superfluid flow if $A<1$ \cite{KaneFisher}. 
For experimentally realistic setups with $v_N/v_J\ll 1$ and the asymptotic behavior 
$A\sim\frac{1}{\sqrt{1-\gamma}}\sqrt{\frac{v_N}{v_J}}$ for $\gamma \rightarrow 1$, 
we find indeed that this transition can only appear for $\gamma \approx 1$. 
This simple criterion provides an upper bound on the transition from the supersolid into the solid. 
However, this transition can be preempted at commensurate fillings with an integer number of particles within each droplet, 
i.e., $n d = 2/\beta \in \mathbb{N}$.
Then, the microscopic interaction between the particles also generates a term \cite{Haldane}
\begin{equation}
  H_{\rs M} = g_{M} \iint dx\, dt \cos\left(2 \vartheta(\bm{x})+ 2 u(\bm{x})/\beta \right),
\end{equation}
which becomes relevant for $D= ({\bf Q}_{11}+{\bf Q}_{12}/\beta)^2+({\bf Q}_{21}+{\bf Q}_{22}/\beta)^2 < 2$ 
and pins the number of particles in each droplet to the integer value $2/\beta$.
It describes the quantum phase transition into a Mott insulator with an excitation gap for adding/removing a particle from a droplet.
However, the droplets can still fluctuate in position giving rise to a phononic sound mode characteristic for a solid.
Since $D=A^{-1}$  for $\gamma\rightarrow 1$, the Mott transition at commensurate fillings occurs earlier than the transition of a  single Josephson Junction.

In conclusion, we have studied the influence of quantum fluctuations on a one-dimensional 
supersolid, and determined the quasi long range order for the off-diagonal as well as the diagonal order. The quantum phase transition from the superfluid to the supersolid is shifted by quantum fluctuations from its mean-field prediction, where the local formation of droplets take place. Furthermore,
the quantum nucleation of phase slips provide a weak dissipation with an algebraic behavior 
between the pressure difference and the particle flow.  However, for current experimental
parameters  for Dysprosium atoms with many atoms per lattice site, these effects are
extremely weak and therefore, the supersolid is accurately described by
mean-field theory.

\acknowledgments  
This work is supported by the German Research Foundation (DFG) within FOR2247 under  Bu2247/1-2.

\bibliography{Citations.bib}

\clearpage
\onecolumngrid
\appendix



\section*{Supplementary material}

# Supplemental Material: Quantum fluctuations in one-dimensional supersolids

Chris Bühler,[1] Tobias Ilg,[1] and Hans Peter Büchler[1]

[1]*Institute for Theoretical Physics III and Center for Integrated Quantum Science and Technology, University of Stuttgart, DE-70550 Stuttgart, Germany*
(Dated: November 30, 2022)

## CONSTRUCTION OF THE TRANSFORMATION MATRIX

In this appendix we derive the transformation which decouples the Hamiltonian in Eq. (3) of the main text into two modes with sound velocity $v_\pm$. This can be achieved if we require that the transformation $\mathbf{Q}$ fulfills the equations

$$\left(\mathbf{Q}^{-1}\right)^T [v_J \mathbf{M}_J] \mathbf{Q}^{-1} = \begin{pmatrix} v_+ & 0 \\ 0 & v_- \end{pmatrix}, \qquad \mathbf{Q}[v_N \mathbf{M_N}] \mathbf{Q^T} = \begin{pmatrix} v_+ & 0 \\ 0 & v_- \end{pmatrix}. \tag{1}$$

From these equations it also follows that $\mathbf{Q}$ has to fulfill

$$\mathbf{Q}[v_N v_J \mathbf{M}_N \mathbf{M}_J]\mathbf{Q}^{-1} = \begin{pmatrix} v_+^2 & 0 \\ 0 & v_-^2 \end{pmatrix}, \tag{2}$$

meaning that it diagonalizes the matrix $\mathbf{M}_N\mathbf{M}_J$ with eigenvalues $v_\pm^2$. We can therefore construct $\mathbf{Q}$ by finding the eigenvectors of $\mathbf{M}_N\mathbf{M}_J$. These are

$$\mathbf{p}_\sigma = \frac{1}{N_\sigma}\left(\tfrac{1}{2}\left[1 - \beta^2\lambda/\gamma + \sigma\sqrt{\Delta}\right], \xi - \beta\lambda\right)^T, \quad \sigma \in \{+, -\}, \tag{3}$$

with $\Delta = (1 - 2\xi\beta + \beta^2\lambda/\gamma)^2 - 4\beta^2(\lambda - \xi^2)(1-\gamma)/\gamma$ and yet undetermined constants $N_\sigma$. The matrix is then defined as $\mathbf{Q}^{-1} = (\mathbf{p}_+ \ \mathbf{p}_-)$. The still undetermined constants $N_\sigma$ must now be used to make sure that Eq. (1) is fulfilled. We obtain that

$$N_\pm^2 = \pm\sqrt{\Delta}\frac{v_\pm^2 - v_N v_J \beta^2 \lambda(1-\gamma)/\gamma}{v_N v_\pm}, \tag{4}$$

which fully determines the transformation. A bit of calculation leads to a somewhat compact form

$$\mathbf{Q}^{-1} = \sqrt[4]{\frac{v_N}{v_J}} \begin{pmatrix} s_{1+}\sqrt{\frac{\gamma \hat{v}_+^2 - \beta^2(\lambda - \xi^2)}{\gamma \hat{v}_+ \sqrt{\Delta}}} & s_{1-}\sqrt{-\frac{\gamma \hat{v}_-^2 - \beta^2(\lambda - \xi^2)}{\gamma \hat{v}_- \sqrt{\Delta}}} \\ s_{2+}\sqrt{\frac{\lambda \hat{v}_+^2 - (\lambda - \xi^2)}{\hat{v}_+ \sqrt{\Delta}}} & s_{2-}\sqrt{-\frac{\lambda \hat{v}_-^2 - (\lambda - \xi^2)}{\hat{v}_- \sqrt{\Delta}}} \end{pmatrix}, \tag{5}$$

with $s_{i\pm} = \text{sign}\left((\mathbf{p}_\pm)_i\right)$ and $\hat{v}_\sigma = v_\sigma/\sqrt{v_J v_N}$. The inverse is then given by

$$\mathbf{Q} = \sqrt[4]{\frac{v_J}{v_N}} \begin{pmatrix} \sqrt{\frac{\gamma \hat{v}_+^2 - \beta^2\lambda(1-\gamma)}{\gamma \hat{v}_+ \sqrt{\Delta}}} & -s_{1-}s_{2-}\beta\sqrt{\frac{\hat{v}_+^2 - (1-\gamma)}{\gamma \hat{v}_+ \sqrt{\Delta}}} \\ -\sqrt{-\frac{\gamma \hat{v}_-^2 - \beta^2\lambda(1-\gamma)}{\gamma \hat{v}_- \sqrt{\Delta}}} & s_{1+}s_{2+}\beta\sqrt{-\frac{\hat{v}_-^2 - (1-\gamma)}{\gamma \hat{v}_- \sqrt{\Delta}}} \end{pmatrix}. \tag{6}$$

## CONNECTION TO MICROSCOPIC QUANTITIES

To estimate the parameters of our model, we consider a reduced three-dimensional model of dipolar bosons of mass $m$ in a transverse harmonic trap of oscillator length $l_\perp$ within mean-field theory and include quantum fluctuations in local-density approximation. For a detailed discussion of this approach we refer the reader to [1]. In the transverse direction we use the ansatz $\psi(y,z) = \frac{1}{\sqrt{\pi}\sigma}e^{-(\nu y^2 + z^2/\nu)/\sigma^2}$ for the wave function, where $\sigma$ and $\nu$ are variational parameters determined by minimizing the ground state energy. In the supersolid regime, we use the mean-field ansatz

$$\phi(x) = \frac{\sqrt{n}}{\sqrt{1 + \sum_{j=1}^\infty \Delta_j^2/2}}\left(1 + \sum_{j=1}^\infty \Delta_j \cos\left[j\, k_\mathrm{s} x\right]\right) \tag{7}$$

for the longitudinal direction, with the order parameters $\Delta_j$, the wave vector $k_s$ of the modulation and the one-dimensional density $n$. We write the energy as

$$E = E_t(\sigma,\nu) - \int dx\, \phi^*(x)\frac{\hbar^2 \nabla^2}{2m}\phi(x) + \frac{1}{2}\int dx\, dx'\, V(x-x')|\phi(x')|^2|\phi(x)|^2 + \frac{2}{5}\tilde{\gamma}\int dx\, |\phi(x)|^5, \qquad (8)$$

where $E_t = \frac{N\hbar^2}{4ml_\perp^2}(\frac{1}{\nu}+\nu)(\frac{1}{\sigma^2}+\sigma^2)$ is the energy contribution of the transverse confinement of the $N$ particles, $V(x)$ is the effective 1D dipolar interaction potential in real space, which we obtain by integrating out the transverse degrees of freedom of the three-dimensional dipolar interaction potential with dipolar strength $\varepsilon_{\rm dd}$ and scattering length $a_s$, assuming the shape $\psi(y,z)$ for the transverse wave function. The last term in Eq. (8) takes into account fluctuations within local-density approximation and $\tilde{\gamma}$ controls the strength of these fluctuations. The ground state is obtained by minimizing Eq. (8) with respect to $\Delta_l$ and $k_s$ as well as $\sigma$ and $\nu$. From the ground state we can extract all relevant quantities of our model. Within mean-field theory, Leggetts upper bound for the superfluid fraction [2] is completely saturated [3],

$$f_{\rm s} = \left[\left(\frac{k_{\rm s}}{2\pi}\int_0^{2\pi/k_{\rm s}} dx\, |\phi(x)|^2\right)\left(\frac{k_{\rm s}}{2\pi}\int_0^{2\pi/k_{\rm s}} dx\, \frac{1}{|\phi(x)|^2}\right)\right]^{-1} \qquad (9)$$

and we obtain $\gamma = 1 - f_s$. The parameters $\kappa$, $\xi'$ and $\lambda'$ of our model characterize the effective potential of the supersolid. By considering the ground sate of Eq. (8) for a fixed chemical potential $\mu$, we can connect variations of the wave vector $k_s$ and the density $n$ to the parameters $\kappa$, $\xi'$ and $\lambda'$. Variations in the density $\delta n$ we can directly connect to $\kappa$, while a variation in the wave vector $k_s$ is connected to $\lambda$, since a linear displacement field $\partial_x u = \delta a = {\rm const.}$ changes the periodicity of the system, $k_s \to k_s - \delta a k_s = k_s + \delta k_s$. We expand $E - \mu N$ up to second order in $\delta n$ and $\delta k_s$,

$$\frac{(E-\mu N)a_{\rm s}}{L} = \frac{\hbar^2}{ml_\perp^2}\left[C_0 + C_\kappa (a_s\delta n)^2 + C_\lambda (l_\perp \delta k_s)^2 + C_\xi (a_s\delta n)(l_\perp \delta k_s)\right], \qquad (10)$$

and we obtain the expansion coefficients $C_\kappa$, $C_\lambda$ and $C_\xi$ numerically. Comparing the Hamiltonian Eq. (3) of the main text to the expansion in Eq. (10) we obtain the dimensionless parameters

$$\lambda = \frac{\pi^2 l_\perp^2 C_\lambda}{a_s^2 C_\kappa}, \qquad \xi = -\frac{\pi l_\perp C_\xi}{2 a_s C_\kappa}, \qquad \beta = \frac{k_s}{n\pi}, \qquad \frac{v_J}{v_N} = \frac{\pi^2 n l_\perp^2}{2 a_s C_\kappa}. \qquad (11)$$

For Fig. (2) of the main text we fix the critical point in mean-field theory to $\varepsilon_{\rm dd,c} = 1.34$, $a_s/l_\perp = 1/200$ while $n \approx 11.931/a_s$ and determine the dimensionless parameters in Eq. (11) while using two order parameters.

--------



\end{document}